\documentclass[12pt,a4paper,epsfig]{article}
\usepackage{color}
\usepackage{amsfonts}
\usepackage{epsfig}

\date{}
\begin{document}

\title{\Huge{\bf Quantum scalar fields in the half-line.
\\ A heat kernel/zeta function approach.}}

\author{ J. Mateos Guilarte$^1$, J. M. Mu$\tilde{\rm n}$oz Casta$\tilde{\rm
n}$eda$^2$, and M. J. Senosiain$^3$
\\ \footnotesize{{ \sl $^1$Departamento de F\'{\i}sica Fundamental and IUFFyM,}}\\ \footnotesize{{\sl Universidad de Salamanca}}\\
\footnotesize{{ \sl $^2$Departamento de F\'{\i}sica Te\'orica}}
\\ \footnotesize{{\sl Universidad de Zaragoza, SPAIN.}}\\
\footnotesize{{\sl $^3$Departamento de Matem\'aticas}} \\ \footnotesize{{\sl Universidad de Salamanca, SPAIN.}}}

\maketitle

\begin{abstract}
In this paper we shall study vacuum fluctuations of a single scalar
field with Dirichlet boundary conditions in a finite but very long
line. The spectral heat kernel, the heat partition function and the
spectral zeta function are calculated in terms of Riemann Theta
functions, the error function, and hypergeometric ${}_PF_Q$
functions.
\end{abstract}

\section{Introduction}
In collaboration with J. Sesma, J. Abad devoted part of the last
years of his fertile scientific career to studying the r$\hat{\rm
o}$le of special functions in quantum field theory. In this brief
memoir, elaborated to honor Julio's memory, we explore the influence
of using Dirichlet boundary conditions in quantum field theory.
Specifically, we shall address the Higgs model in (1+1)-dimensions
but we shall restrict the spatial line to become a finite interval.
Then, Dirichlet boundary conditions at the endpoints of the interval
will be imposed on the field. Eventually, we shall allow the length
of the interval to tend to infinity to describe the situation in
which the mesons meet an impenetrable wall. Our playground is thus
the analysis of scalar quantum fields living in a half-line.

In this short work we shall concentrate on computing very basic
quantities. Essentially, we shall deal with vacuum fluctuations in
such a way that the spectral zeta function of the second-order
differential operator governing small fluctuations around the vacuum
will be used to regularize the divergent zero-point energy. The
spectral information is also encoded in the associated $K$-heat
partition function and $K$-heat kernel. These spectral functions
permit a high-temperature asymptotic expansion, which, in turn,
determines via the Mellin transform the meromorphic structure of the
spectral zeta function in terms of the heat coefficients. The main
sources of our approach are References \cite{Eli}, \cite{Kir}, and
\cite{Vass} as well as \cite{mdj1} and \cite{mdj2}. We hope that
Julio would have been pleased with our results. In recent times he
was one of those rare theorists trusted and praised by experimental
and applied physicists.

\section{The Higgs model in a line}

In the $(1+1)$-dimensional toy Higgs model the action
\[
S=\int \, dy^2 \, \left\{{1\over 2}\frac{\partial\psi}{\partial
y^\mu}\frac{\partial\psi}{\partial y_\mu}-{\lambda\over
4}(\psi^2(y_0,y)-{m^2\over\lambda})^2\right\}
\]
governs the dynamics of the scalar field $\psi(y_0,y): {\mathbb
R}^{1,1}\rightarrow {\mathbb R}$. We choose the metric
$g_{\mu\nu}={\rm diag}(1,-1)$ in (1+1)-dimensional ${\mathbb
R}^{1,1}$ Minkowskian space-time. In the natural system of units
$\hbar=c=1$ the dimension of the field, the mass, and the coupling
constant are respectively: $[\psi]=1$, $[\lambda]=[m^2]=L^{-2}$. In
terms of non-dimensional space-time coordinates and fields
\[
y^\mu\rightarrow y^\mu={\sqrt{2}\over m}\cdot x^\mu \hspace{1.5cm} ;
\hspace{1.5cm} \psi(y^\mu)\rightarrow
\psi(y^\mu)={m\over\sqrt{\lambda}}\cdot \phi(x^\mu) \qquad ,
\]
the action functional and the field equations of the
$\lambda(\phi)^4_2$ model read:
\[
S={m^2\over\lambda}\int \, dx^2 \, \left\{{1\over
2}\frac{\partial\phi}{\partial x^\mu}\frac{\partial\phi}{\partial
x_\mu}-{1\over 2}(\phi^2(x_0,x)-1)^2\right\}
\]
\[
\frac{\partial^2\phi}{\partial
x_0^2}(x_0,x)-\frac{\partial^2\phi}{\partial
x^2}(x_0,x)=2\phi(x_0,x)(1-\phi^2(x_0,x)) \qquad .
\]
The shift of the scalar field from the homogeneous stable solution,
$\phi(x^\mu)=1+H(x^\mu)$, leads to the action
\[
S=\frac{m^2}{\lambda}\int \, d^2x \, \left\{\left[{1\over
2}\partial_\mu H\partial^\mu H -2H^2(x^\mu)
\right]-\left[2H^3(x^\mu)+{1\over 2}H^4(x^\mu)\right]\right\} \qquad
,
\]
which shows the spontaneous symmetry breakdown of the internal
parity ${\mathbb Z}_2$ symmetry.

\section{{\bf Zero point vacuum energy with Dirichlet boundary
conditions}}

The linearized field equations
\begin{equation}
\frac{\partial^2\delta H}{\partial
x_0^2}(x_0,x)-\frac{\partial^2\delta H}{\partial x^2}(x_0,x)+4\delta
H(x_0,x)=0 \label{lfe}
\end{equation}
allow us to expand the Higgs field $\delta H(x_0,x)$ as a linear
superposition of solutions obtained by means of separation of
variables:
\begin{equation}
\delta
H(x_0,x)=\frac{1}{m}\cdot\sqrt{\frac{\lambda}{l}}\sum_k{1\over\sqrt{2\omega(k)}}
\left\{a(k)e^{-ik_0x_0}f(x;k)+a^*(k)e^{ik_0x_0}f^*(x;k)\right\}
\qquad . \label{glfe}
\end{equation}
(\ref{glfe}) is the general solution of (\ref{lfe}) if the
dispersion relation between the frequency and energy of the plane
waves $k_0^2-k^2-4=0$ ($k_0=\omega(k)=\sqrt{k^2+4}$) holds. Of
course, $f(x;k)$ are the eigenfunctions of the second-order
fluctuation operator:
\begin{equation}
K_0=-{d^2\over dx^2}+4 \qquad \qquad , \qquad
K_0f(x;k)=\omega^2(k)f(x;k) \label{sofo} \qquad .
\end{equation}

In the normalization interval $I=[0,l\,]$, $l={mL\over\sqrt{2}}$,
the spectrum of $K_0$ with Dirichlet boundary conditions (following the method developed in \cite{mj})
\[
K_0 f_n(x)=\omega_n^2 f_n(x) \qquad , \qquad f_n(0)=f_n(l)=0 \qquad
\]
is:
\[
k_n=\frac{\pi}{l}n \qquad , \qquad \omega_n^2={\pi^2\over l^2}n^2+4
\qquad , \qquad f_n(x)=\sqrt{{2\over l}}\,\,{\rm sin}({\pi\over l} n
x) \qquad , \qquad n\in{\mathbb Z}^+ \qquad .
\]
Therefore, the classical Hamiltonian is tantamount to an infinite
number of oscillators given by the Fourier coefficients of these
standing waves:
\begin{eqnarray*}
H^{(2)}&=&{m^3\over\sqrt{2}\lambda}\int \, dx \, \left\{{1\over
2}\frac{\partial\delta H}{\partial x_0}\cdot\frac{\partial\delta H
}{\partial x_0}+{1\over 2}\frac{\partial\delta H}{\partial
x}\cdot\frac{\partial\delta H}{\partial x}+\delta
H(x_0,x)\cdot\delta H(x_0,x) \right\}\\&=&{m\over
2\sqrt{2}}\sum_{n=1}^\infty \,
\omega(k_n)\left(\frac{}{}a^*(k_n)a(k_n)+a(k_n)a^*(k_n)\right)
\qquad .
\end{eqnarray*}
Canonical quantization $[{\hat a}(k_n),{\hat
a}^\dagger(k_m)]=\delta_{nm}$ promotes the Fourier coefficients to
creation and annihilation operators and gives the free quantum
Hamiltonian:
\[
\hat{H}^{(2)}_0= {m\over\sqrt{2}}\sum_{n=1}^\infty \,
\omega(k_n)\left(\hat{a}^\dagger(k_n)\hat{a}(k_n)+{1\over 2}\right)
\qquad .
\]
It is clear that the vacuum $\left(\hat{a}^\dagger(k_n)|0>=0,
\forall n\right)$ energy is not zero but:
\[
\Delta E_0=<0|\hat{H}|0>={ m\over 2\sqrt{2}}\sum_{n=1}^\infty
\omega(k_n)={ m\over 2\sqrt{2}}{\rm Tr}_D K_0^{{1\over 2}}
\hspace{1cm} ,
\]
a divergent quantity.
\subsection{{\bf The heat function}}

Better expectations of convergence are offered by another spectral
function, the  $K_0$-heat function:
\begin{equation}
{\rm Tr}_D e^{-\beta K_0}=\int_0^l \, dx \,
K_{K_0}(x,x;\beta)=\sum_{n=1}^\infty \,
e^{-\beta\left(\frac{\pi^2n^2}{l^2}+4\right)} \label{hpf}
\end{equation}
where $K_{K_0}(x,y;\beta)$ is the kernel of the $K_0$-heat equation
\[
\left(\frac{\partial}{\partial\beta}+K_0\right)\Psi(\beta,x)=\left(\frac{\partial}{\partial\beta}-{\partial^2\over
\partial x^2}+4\right)K_{K_0}(x,y;\beta)=0 \qquad , \qquad
K_{K_0}(x,y;0)=\delta(x-y)
\]
and $\beta=\frac{m}{k_B T}$ is proportional to the inverse
temperature. Moreover, via the Mellin transform the spectral zeta
function is obtained:
\begin{equation}
\zeta_{K_0}(s)=\frac{1}{\Gamma(s)}\int_0^\infty \, d\beta \,
\beta^{s-1}{\rm Tr}_D e^{-\beta K_0}=\sum_{n=1}^\infty\,
\frac{1}{(\frac{\pi^2 n^2}{l^2}+4)^s} \label{dsf} \qquad .
\end{equation}
We shall use this meromorphic function of the complex variable $s$
(and will return to this later) to regularize the divergent sum of
vacuum fluctuations, $\Delta E_0$, by assigning to it the value of
the series at a regular point in the $s$ complex plane.
\subsubsection{{\bf Riemann Theta constants}}
The $K_0$-heat function is essentially given by a Riemann Theta
constant:
\begin{equation}
{\rm Tr}_D\, e^{-\beta \, K_0}=\sum_{n=1}^\infty\, e^{-\beta \,
\omega_n^2}=\frac{e^{-4\beta}}{2}\left(\sum_{n=-\infty}^\infty\,
{\rm exp}[-\beta\, \frac{\pi^2}{l^2}\,
n^2]-1\right)=\frac{e^{-4\beta}}{2}\left(\Theta\left[\begin{array}{c}
0
\\ 0 \end{array}\right](0|i\, \frac{\pi}{l^2}\, \beta)-1\right)
\, \,  . \label{dhf}
\end{equation}
Here, we denote the very well known Riemann or Jacobi Theta
functions in the form:
\[
\Theta\left[\begin{array}{c}a \\
b\end{array}\right](z|\tau)=\sum_{n=-\infty}^\infty \, {\rm
exp}\left[2\pi i[(n+a)(z+b)+{1\over 2}(n+a)^2\tau]\right] \qquad ,
\]
$z\in{\mathbb C}\, \, \, , \, \, \, \tau\in{\mathbb C}, {\rm
Im}\tau>0 \, \, \, , \, \, \, a,b=0,{1\over 2}$. Thus, we need the
Riemann Theta function at the $z=0$ point (Theta constant), the
modular parameter $\tau=i\frac{\pi}{l^2}\, \beta$ (determined by
$\beta$ and $l$), and the \lq\lq characteristics" $a=b=0$. Use of
the Poisson formula
\[
\Theta\left[\begin{array}{c} 0 \\ 0 \end{array}\right](0, i\,
\frac{\pi}{l^2}\,
\beta)=\frac{l}{\sqrt{\pi\beta}}\,\,\Theta\left[\begin{array}{c} 0 \\
0
\end{array}\right](0, i\, \frac{l^2}{\pi\beta})
\]
allows us to write the $K_0$-heat function in the new form:
\[
{\rm Tr}_D\, e^{-\beta \,
K_0}=\frac{e^{-4\beta}}{2}\left(\frac{l}{\sqrt{\pi\beta}}\Theta\left[\begin{array}{c}
0 \\ 0
\end{array}\right](0, i\, \frac{l^2}{\pi\beta})-1\right) \qquad .
\]
From this, an asymptotic formula for the behavior of the $K_0$-heat
function is obtained:
\begin{equation}
\Theta\left[\begin{array}{c} 0 \\ 0
\end{array}\right](0, i\, \frac{l^2}{\pi\beta})\cong 1+{\cal
O}(e^{-{c\over\beta}}) \quad \Rightarrow \quad {\rm Tr}_D\,
e^{-\beta \, K_0}\cong_{\beta\rightarrow
0}\frac{e^{-4\beta}}{2}\left(\frac{l}{\sqrt{\pi\beta}}-1\right)+{\cal
O}(e^{-{c\over\beta}}) \qquad . \label{adhf}
\end{equation}

\subsubsection{{\bf Physicists' derivation: the Error function}}
We now offer a derivation of the asymptotic formula by means of
physicists' techniques. The idea is to look at the problem when $l$
is very large: $l\,\rightarrow \, \infty$. The spectral density of
the standing waves can be determined from the phase shifts
$\delta^D(k)=-{\rm Si}(2 k l)$ (${\rm Si}(x)$ is the sine integral
function) due to the reflected waves:
\[
{\rm sin}\left(k l+\delta^D(k)\right)=0 \qquad \equiv \qquad k l +
\delta^D(k)=n \pi \quad , \quad n\in{\mathbb Z}^+
\]
\[
\rho_{K_0}^D(k)=\frac{d n}{d k}={l\over\pi}+{1\over
\pi}.\frac{d\delta^D}{d k}(k)={l\over\pi}\left(1-\frac{{\rm sin}(2 k
l)}{k l}\right) \qquad .
\]
Thus, we end with an integral, rather than a series, for the
$K_0$-heat function in terms of the error function:
\begin{equation}
{\rm Tr}_D\, e^{-\beta\, K_0}={l\over \pi}\int_0^\infty \, dk \,
\left(1-\frac{{\rm sin}2 k l}{k l}\right)\cdot e^{-\beta(k^2+4)}=
{e^{-4\beta}\over 2}\left({l\over\sqrt{\pi\beta}}-{\rm
Erf}\left[{l\over\sqrt{\beta}}\right]\right) \qquad . \label{sdhf}
\end{equation}
The high-temperature formula agrees  perfectly with (\ref{adhf})
\[
{\rm Erf}[{l\over\sqrt{\beta}}]\cong_{\beta\rightarrow 0}\, 1+{\cal
O}(e^{-{c\over\beta}}) \quad \Rightarrow \quad {\rm Tr}_D\,
e^{-\beta \, K_0}\cong_{\beta\rightarrow
0}\frac{e^{-4\beta}}{2}\left(\frac{l}{\sqrt{\pi\beta}}-1\right)+{\cal
O}(e^{-{c\over\beta}})
\]
and, neglecting exponentially small contributions, we find the
coefficients of the high-temperature expansion:
\begin{eqnarray*}
{\rm Tr}_D\,e^{-\beta\,
K_0}&=&e^{-4\beta}\cdot\left(\frac{l}{\sqrt{4\pi\beta}}-{1\over
2}{\rm Erf}[\frac{l}{\sqrt{\beta}}]\right)=e^{-4\beta}\cdot\sum_n\,
c_n(K_0)\, \beta^{n-{1\over 2}}  \quad ,  \quad n\in
\{0\}\cup{\mathbb Z}^+_{1/2} \\ &\cong& e^{-4\beta}\left(
{l\over\sqrt{4\pi}}-{1\over 2}+{\cal
O}(e^{-{c\over\beta}})\right)\qquad , \qquad
c_0(K_0)=\frac{l}{\sqrt{4\pi}} \, \,  , \quad c_{1/2}(K_0)=-{1\over
2} \qquad ,
\end{eqnarray*}
$c_n(K_0)=0 \quad , \quad \forall \, n\geq 1 \, \,$.

\subsection{{\bf The spectral zeta function}}
\subsubsection{\bf Epstein zeta function}
Mellin's transform of the $K_0$-heat function (\ref{dhf}) provides
the spectral zeta function in terms of the Epstein zeta function
$E(s,a|A)=\sum_{n=-\infty}^\infty \, \frac{1}{(An^2+a)^s}$:
\begin{eqnarray*}
\zeta_{K_0}^D(s)&=&{1\over\Gamma(s)}\int_0^\infty \, d\beta \,
\beta^{s-1}\, {\rm Tr}_D\,e^{-\beta K_0}={1\over 2\Gamma(s)}\cdot
\int_0^\infty \, d\beta \, \beta^{s-1}\,
\left(\sum_{n=-\infty}^\infty\, e^{-\beta({\pi^2\over
l^2}n^2+4)}-e^{-4\beta}\right)\\&=&{1\over
2}\sum_{n=-\infty}^\infty\,\frac{1}{({\pi^2\over
l^2}n^2+4)^s}-\frac{1}{2^{2s+1}}={1\over 2}\, E(s,4|{\pi^2\over
l^2})-\frac{1}{2^{2s+1}} \qquad .
\end{eqnarray*}
Mellin's transform, however, of the Poisson inverted version
\begin{eqnarray}
\zeta_{K_0}^D(s)&=&{1\over 2\Gamma(s)}\cdot \int_0^\infty \, d\beta
\, \beta^{s-1}\, e^{-4\beta}\,
\left({l\over\sqrt{\pi}}\beta^{-{1\over 2}}\sum_{n=-\infty}^\infty\,
e^{-{l^2\over
\beta}n^2}-1\right)\nonumber\\&=&\frac{1}{\sqrt{\pi}}\cdot\frac{\Gamma(s-1/2)}{4^s\Gamma(s)}
+\frac{l}{2^{s-1/2}\Gamma(s)\sqrt{\pi}}\, \sum_{n\in{\mathbb
Z}/\{0\}} \, (l n)^{s-1/2}\, K_{1/2-s}(4 l n)-
\frac{1}{2^{2s+1}}\quad . \label{dzfk}
\end{eqnarray}
gives the spectral zeta function as a series of modified Bessel
functions of the second type. Moreover, formula (\ref{dzfk}) shows
that there are poles of $\zeta_{K_0}^D(s)$ at the points
\[
s=\frac{1}{2}, -\frac{1}{2}, -\frac{3}{2}, -\frac{5}{2},
-\frac{7}{2}, \cdots, -\frac{2j+1}{2}, \cdots, \qquad j\in{\mathbb
Z}^+
\]
because $K_{1/2-s}(4 l n)$ are transcendental entire functions, i.e.
holomorhic functions of $s$ in ${\mathbb C}/{\infty}$ with an
essential singularity at $s=\infty$.

\subsubsection{\bf Physicists' derivation: Hypergeometric ${}_PF_Q$
functions} Mellin's transform of the (\ref{sdhf}) version of the
$K_0$-heat function
\begin{eqnarray}
\zeta_{K_0}^D(s)&=&{1\over \Gamma(s)}\cdot \int_0^\infty \, d\beta
\, \beta^{s-1}\, {e^{-4\beta}\over
2}\left({l\over\sqrt{\pi\beta}}-{\rm
Erf}\left[{l\over\sqrt{\beta}}\right]\right)\nonumber\\&=&{l\over\sqrt{4\pi}}\cdot\frac{\Gamma(s-1/2)}{\Gamma(s)}
\left({1\over 2^{2s-1}}-{1\over
2^{2(s-1)}}\cdot{}_1F_2[1/2;3/2,3/2-s;4 l^2]\right.\nonumber
\\ &-&\left.
\frac{l^{2s-1}}{s}\cdot{}_1F_2[s;1/2+s,1+s;4 l^2]\right)\quad ,
\label{dzfk1}
\end{eqnarray}
supplies a third analytical expression of the spectral zeta
function. Euler $\Gamma$ functions and hypergeometric ${}_PF_Q$
functions, with power expansion around $z=0$
\[
{}_PF_Q[a_1,a_2, \cdots , a_p;b_1, b_2, \cdots ,b_q;z]=
\sum_{k=0}^\infty \frac{(a_1)_k(a_2)_k \cdots
(a_p)_k}{(b_1)_k(b_2)_k \cdots (b_q)_k}\cdot\frac{z^k}{k!} \qquad ,
\]
where $(a)_k=a(a+1)(a+2)\cdots (a+k-1)$ is the Pochhammer symbol,
enter the third formula of $\zeta_{K_0}^D(s)$. It is clear that the
physical point $s=-{1\over 2}$ is a pole of at least
$\Gamma(s-{1\over 2})$. Other poles come from the other poles of
$\Gamma(s-{1\over 2})$, $s-{1\over 2}=0, -2, -3, -4, -5, \cdots$,
and the poles of ${}_1F_2[1/2;3/2,3/2-s;4 l^2]$ and
${}_1F_2[s;1/2+s,1+s;4 l^2]$, which are meromorphic functions of
$s$. From the residue representation of these functions
\begin{eqnarray*}
{}_1F_2[\frac{1}{2};\frac{3}{2},\frac{3}{2}-s;4l^2]&=&
\frac{\Gamma(\frac{3}{2})\Gamma(\frac{3}{2}-s)}{\Gamma(\frac{1}{2})}\cdot
\sum_{j=0}^\infty \, {\rm res}_u\left(\frac{\Gamma(\frac{1}{2}-u)(-4l^2)^u}{\Gamma(\frac{3}{2}-u)
\Gamma(\frac{3}{2}-s-u)}\Gamma(u)\right)(-j)\\
{}_1F_2[s;\frac{1}{2}+s,1+s;4l^2]&=&
\frac{\Gamma(\frac{1}{2}+s)\Gamma(1+s)}{\Gamma(s)}\cdot
\sum_{j=0}^\infty \, {\rm res}_u\left(\frac{\Gamma(s-u)(-4l^2)^u}{\Gamma(\frac{1}{2}+s-u)\Gamma(1+s-u)}
\Gamma(u)\right)(-j)\\
\end{eqnarray*}
we find poles when $3/2-s=-k_1, 1/2+s=-k_2, 1+s=-k_3,
k_1,k_2,k_3\in{\mathbb Z}^+\bigcup \{0\}$. All together, there are
poles of $\zeta^D_{K_0}(s)$ at:
\[
s=\cdots -5/2,-2,-3/2,-1,-1/2,1/2,3/2,5/2,7/2, \cdots \qquad .
\]

\subsection{\bf The heat equation kernel}
Finally, in this sub-Section we analyze how the $K_0$-heat function,
henceforth the spectral zeta function, are obtained from the
$K_0$-heat kernel.
\subsubsection{\bf Jacobi Theta functions}
The $K_0$-heat equation kernel satisfying the Dirichlet boundary
conditions
\begin{equation}
\left(\frac{\partial}{\partial\beta}-{\partial^2\over
\partial x^2}+4\right)K_{K_0}^D(x,y;\beta)=0 \, \, \, , \, \, \,
K_{K_0}^D(x,y;0)=\delta(x-y) \quad , \quad
K_{K_0}^D(0,y;\beta)=K_{K_0}^D(l,y;\beta)=0 \, . \label{dhek}
\end{equation}
is:
\begin{eqnarray}
K_{K_0}^D(x,y;\beta)&=&{2\over l}e^{-4 \beta}\sum_{n=1}^\infty\,
{\rm sin}({\pi\over l}n x){\rm sin}({\pi\over l}n y)\cdot
e^{-\beta{\pi^2\over l^2}n^2}={e^{-4 \beta}\over
l}\sum_{n=-\infty}^\infty\, {\rm sin}({\pi\over l}n x){\rm
sin}({\pi\over l}n y)\cdot e^{-\beta{\pi^2\over
l^2}n^2}\nonumber\\&=&{e^{-4 \beta}\over
2l}\sum_{n=-\infty}^\infty\,\left({\rm cos}({\pi\over l}n(x-y))-{\rm
cos}({\pi\over l}n(x+y))\right)\cdot e^{-\beta{\pi^2\over
l^2}n^2}\nonumber\\&=&{e^{-4 \beta}\over
2l}\cdot\left(\Theta\left[\begin{array}{c} 0
\\ 0 \end{array}\right](\frac{x-y}{2 l}|i\frac{\pi}{l^2}\beta)-\Theta\left[\begin{array}{c} 0
\\ 0 \end{array}\right](\frac{x+y}{2
l}|i\frac{\pi}{l^2}\beta)\right) \qquad . \label{tdhek}
\end{eqnarray}
Alternatively, a modular transformation allows us to express the
heat kernel in the new form:
\begin{eqnarray*}
K_{K_0}^D(x,y;\beta)&=&e^{-4 \beta}
\cdot{1\over\sqrt{4\pi\beta}}\cdot\left(e^{-{(x-y)^2\over
4\beta}}\Theta\left[\begin{array}{c} 0
\\ 0 \end{array}\right](-i l \frac{x-y}{2 \beta}|i\frac{l^2}{\pi\beta})\right.\\
&-& \left. e^{-{(x+y)^2\over 4\beta}}\Theta\left[\begin{array}{c} 0
\\ 0 \end{array}\right](-i l\frac{x+y}{2
\beta}|i\frac{l^2}{\pi\beta})\right)\qquad ,
\end{eqnarray*}
because the Jacobi theta functions involved are modular forms of
weight $1/2$.

\subsubsection{\bf Physicists' derivation: the Laplace transform}

Another route to solve (\ref{dhek}) is to look for solutions of the
form
\begin{equation}
K_{K_0}^D(x,y;\beta)=K_{K_0}(x,y;\beta)+e^{-4\beta}D(x,y;\beta)
\label{dheka}
\end{equation}
where
\[
K_{K_0}(x,y;\beta)=\frac{e^{-4\beta}}{\sqrt{4\pi\beta}}\cdot{\rm
exp}[-\frac{(x-y)^2}{4\beta}]
\]
is the $K_0$-heat equation kernel with periodic boundary conditions.
(\ref{dheka}) complies with Dirichlet boundary conditions if:
\begin{equation}
D(x,y;0)=0 \quad , \quad
D(0,y;\beta)=-\frac{1}{\sqrt{4\pi\beta}}\cdot
e^{-\frac{y^2}{4\beta}} \quad , \quad
D(l,y;\beta)=-\frac{1}{\sqrt{4\pi\beta}}\cdot
e^{-\frac{(l-y)^2}{4\beta}} \, \,  . \label{ddhek}
\end{equation}
The Dirichlet boundary conditions (\ref{ddhek}) force the Laplace
transform of $D(x,y;\beta)$, $\bar{D}(x,y;s)=\int \, d\beta \,
e^{-s\beta}D(x,y;\beta)$, to satisfy:
\begin{equation}
\bar{D}(0,y;s)=-\frac{e^{-\sqrt{s}y}}{2\sqrt{s}} \qquad , \qquad
\bar{D}(l,y;s)=-\frac{e^{-\sqrt{s}(l-y)}}{2\sqrt{s}} \qquad .
\label{lddhek}
\end{equation}
Moreover, the ansatz (\ref{dheka}) solves (\ref{dhek}) if
$\bar{D}(x,y;s)$ solves the Laplace equation:
\begin{equation}
\left(\frac{d^2}{d x^2}-s\right)\bar{D}(x,y;s)=0 \qquad . \label{le}
\end{equation}
The general solution of (\ref{le}) is
\[
\bar{D}(x,y;s)=A(y)e^{-\sqrt{s}x}+B(y)e^{\sqrt{s}x}
\]
which complies with (\ref{lddhek}) if:
\begin{eqnarray}
\bar{D}(x,y;s)={1\over 2\sqrt{s}}&\cdot&\left[\frac{{\rm
exp}[-\sqrt{s}(l+x-y)]-{\rm
exp}[-\sqrt{s}(x+y-l)]}{e^{l\sqrt{s}}-e^{-l\sqrt{s}}}\right.\nonumber\\&+&\left.\frac{{\rm
exp}[-\sqrt{s}(l-x+y)]-{\rm
exp}[-\sqrt{s}(l-x-y)]}{e^{l\sqrt{s}}-e^{-l\sqrt{s}}}\right] \qquad
. \label{ltdhk}
\end{eqnarray}

The last step is to take the inverse Laplace transform of
$\bar{D}(x,y;s)$ as given in (\ref{ltdhk}). To do this, it is
convenient to write the common denominator as a power series
expansion:
\[
\frac{1}{e^{l\sqrt{s}}-e^{-l\sqrt{s}}}=\frac{e^{-l\sqrt{s}}}{1-e^{-2l\sqrt{s}}}=\sum_{n=0}^\infty
\, (-1)^n e^{-(2n+1)l\sqrt{s}} \qquad ,
\]
or,
\begin{eqnarray*}
\bar{D}(x,y;s)=\frac{1}{2\sqrt{s}}&\cdot&\sum_{n=0}^\infty\, (-1)^n
\, \left[{\rm exp}[-\sqrt{s}(2l(n+1)+x-y)]-{\rm
exp}[-\sqrt{s}(2nl+x+y)]\right.\\&+&\left.{\rm
exp}[-\sqrt{s}(2l(n+1)-x+y)]-{\rm
exp}[-\sqrt{s}(2l(n+1)-x-y)]\right] \qquad .
\end{eqnarray*}
The inverse Laplace transform of this is easy and gives:
\begin{eqnarray*}
D(x,y;\beta)=\frac{1}{\sqrt{4\pi\beta}}&\cdot&\sum_{n=0}^\infty\,
(-1)^n \, \left[{\rm exp}[-\frac{(2l(n+1)+x-y)^2}{4\beta}]-{\rm
exp}[-\frac{(2ln+x+y)^2}{4\beta}]\right.\\&+&\left.{\rm
exp}[-\frac{(2l(n+1)-x+y)^2}{4\beta}]-{\rm
exp}[-\frac{(2l(n+1)-x-y)^2}{4\beta}]\right] \qquad .
\end{eqnarray*}
From this formula we derive the Dirichlet $K_0$-heat kernel at
coinciding points
\[
K_{K_0}^D(x,x;\beta)={e^{-4\beta}\over\sqrt{4\pi\beta}}\cdot\left[1+\sum_{n=0}^\infty\,
(-1)^n\left(2e^{-\frac{l^2(n+1)^2}{\beta}}-e^{-\frac{(ln+x)^2}{\beta}}-e^{-\frac{(l(n+1)-x)^2}{\beta}}\right)\right]
\]
which in turn provide the $K_0$ heat function through integration on
the interval:
\begin{eqnarray}
{\rm Tr}_De^{-\beta K_0}&=&\int_0^l \, dx \, K_{K_0}^D(x,x;\beta)={l
e^{-4\beta}\over\sqrt{4\pi\beta}}\left(1+2\sum_{n=0}^\infty\,
(-1)^n\,
e^{-\frac{l^2(n+1)^2}{\beta}}\right)\nonumber\\&-&{e^{-4\beta}\over\sqrt{4\pi\beta}}\cdot\sum_{n=0}^\infty\,
(-1)^n\, \int_0^l\, dx \,
\left\{e^{-\frac{(ln+x)^2}{\beta}}+e^{-\frac{(l(n+1)-x)^2}{\beta}}\right\}
\label{ddrhf}\\&=& {l
e^{-4\beta}\over\sqrt{4\pi\beta}}\left(2-\Theta\left[\begin{array}{c}
0 \nonumber\\ 1/2
\end{array}\right](0|i\frac{l^2}{\pi\beta})\right)\\&-& {e^{-4\beta}\over
2}\sum_{n=0}^\infty\, (-1)^n \, \left({\rm
Erf}\left[\frac{l(n+1)}{\sqrt{\beta}}\right]-{\rm Erf}\left[\frac{l
n}{\sqrt{\beta}}\right]\right)\, . \nonumber
\end{eqnarray}
Because
\[
\sum_{n=0}^\infty\, (-1)^n \, \left({\rm
Erf}\left[\frac{l(n+1)}{\sqrt{\beta}}\right]-{\rm Erf}\left[\frac{l
n}{\sqrt{\beta}}\right]\right)\cong_{\beta\to 0} 1+{\cal
O}(e^{-\frac{c}{\beta}})
\]
we again find
\[
{\rm Tr}_D\, e^{-\beta \, K_0}\cong_{\beta\rightarrow
0}\frac{e^{-4\beta}}{2}\left(\frac{l}{\sqrt{\pi\beta}}-1\right)+{\cal
O}(e^{-{c\over\beta}})
\]
in the high-temperature regime.
\section{Summary and outlook}
In sum, we have found three different expressions for the
$K_0$-heat function:
\[
{\rm Tr}_De^{-\beta
K_0}=\frac{e^{-4\beta}}{2}f_1(\frac{\pi}{l^2}\beta)=\frac{e^{-4\beta}}{2}f_2(\frac{\pi}{l^2}\beta)
=\frac{e^{-4\beta}}{2}f_3(\frac{\pi}{l^2}\beta)
\]
where
\begin{eqnarray*}
f_1(|\tau|) &=& \Theta \left[\begin{array}{c} 0 \\ 0
\end{array}\right](0|\tau)-1 \qquad , \qquad |\tau|=\frac{\pi}{l^2}\beta
\\ f_2(|\tau|)&=&\frac{1}{\sqrt{|\tau|}}-{\rm
Erf}[\sqrt{\frac{\pi}{|\tau|}}]
\\ f_3(|\tau|)&=&\frac{2}{\sqrt{|\tau|}}\left(1-{1\over
2}\Theta\left[\begin{array}{c}0 \\ 1/2
\end{array}\right](0|-\frac{1}{\tau})\right)-\sum_{n=0}^\infty(-1)^n\left({\rm Erf}\left[(n+1)\sqrt{\frac{\pi}{|\tau|}}\right]
-{\rm Erf}\left[n\sqrt{\frac{\pi}{|\tau|}}\right]\right) \quad .
\end{eqnarray*}

\begin{figure}[htbp]
\centerline{\includegraphics[height=3.5cm]{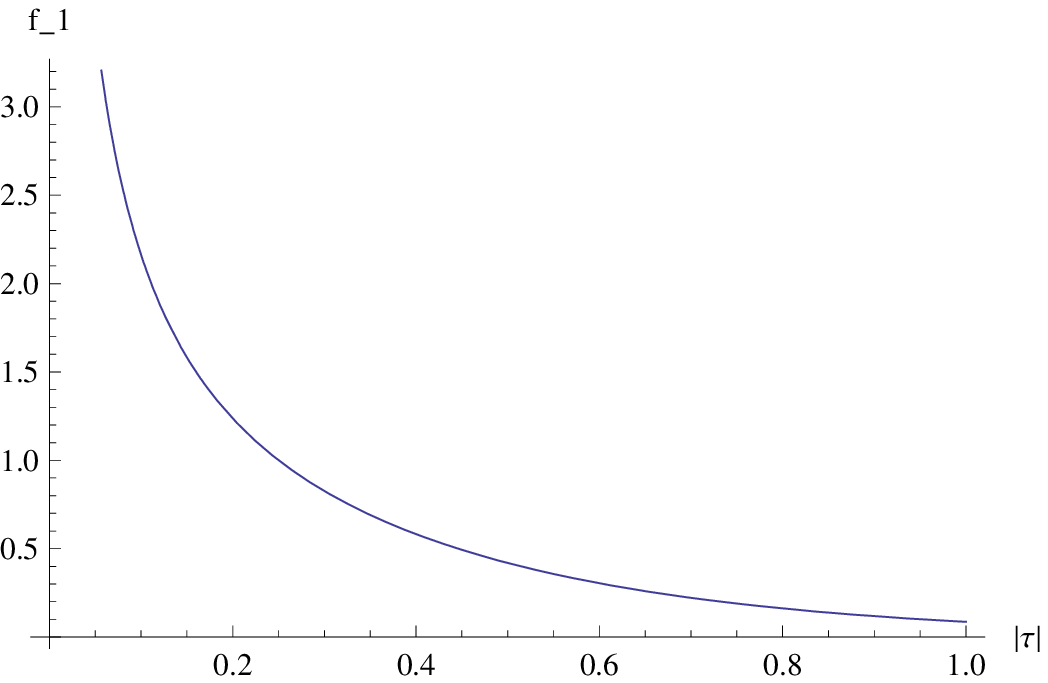}\quad \quad
\includegraphics[height=3.5cm]{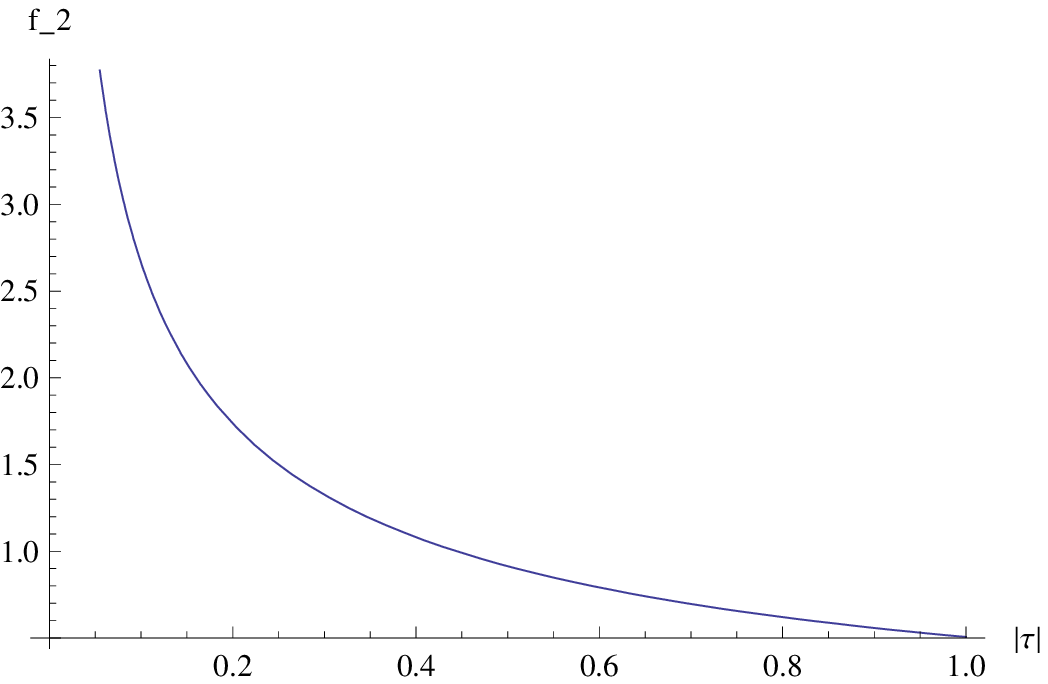}\quad \quad
\includegraphics[height=3.5cm]{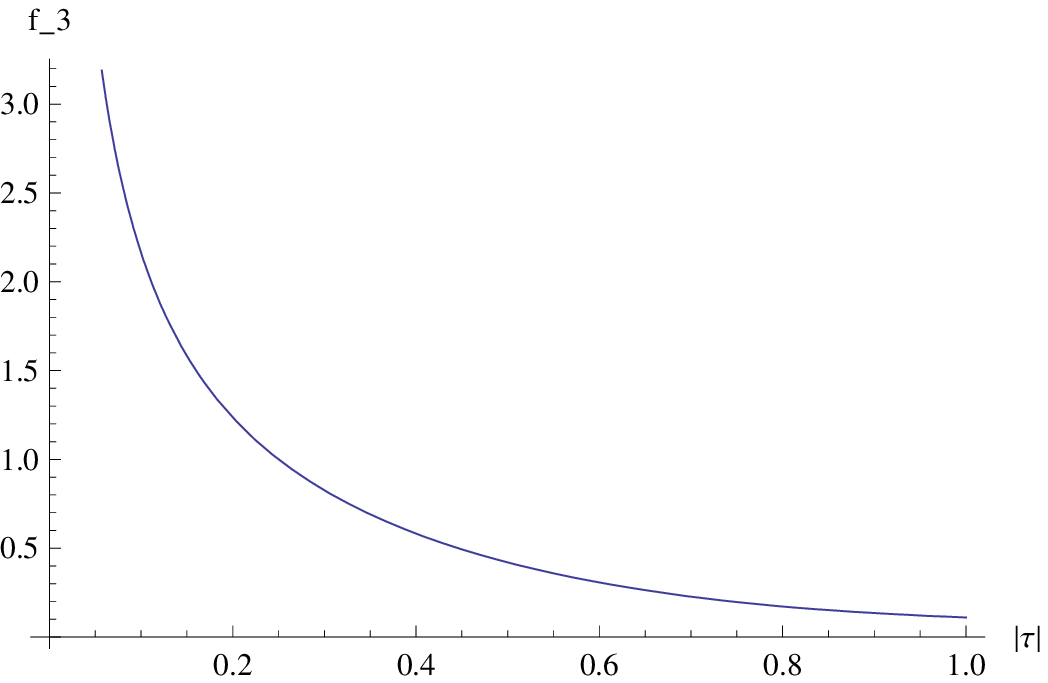}} \caption {
Plot of: a) $f_1(|\tau|)$, b)$f_2(|\tau|)$, and c) $f_3(|\tau|)$.}
\end{figure}
Figure 1 shows the Mathematica graphics of $f_1(|\tau|)$,
$f_2(|\tau|)$ and $f_3(|\tau|)$. In Figure 2(a) the graphics of
$f_1(|\tau|)$ and $f_2(|\tau|)$ are shown together. Simili modo, the
graphics of $f_1(|\tau|$ and $f_3(|\tau|)$ are plotted together in
Figure 2(b). It is clear that all three graphics agree perfectly
when $\beta\to 0$ (high-temperature) and/or $l\to\infty$ (infinite
length of the interval). $f_1(|\tau|)$ and $f_2(|\tau|)$, however,
start to differ at $|\tau|=0.7$, whereas there are no differences in
the graphics of $f_1(|\tau|)$ and $f_3(|\tau|)$. It is amazing how
two different derivations involving highly sophisticated special
functions lead to identical curves ! From a physical point of view
we are tempted to speculate that $f_3(|\tau|)$ would give the exact
result $f_1(|\tau|)$ because the infinite rebounds of the standing
waves in the walls at $x=0$ and $x=l$ are accounted for. Instead,
$f_2(x)$ counts a single rebound in the $x=0$ wall, which is a
legitimate approximation for $l\to\infty$.
\begin{figure}[htbp]
\centerline{\includegraphics[height=4.5cm]{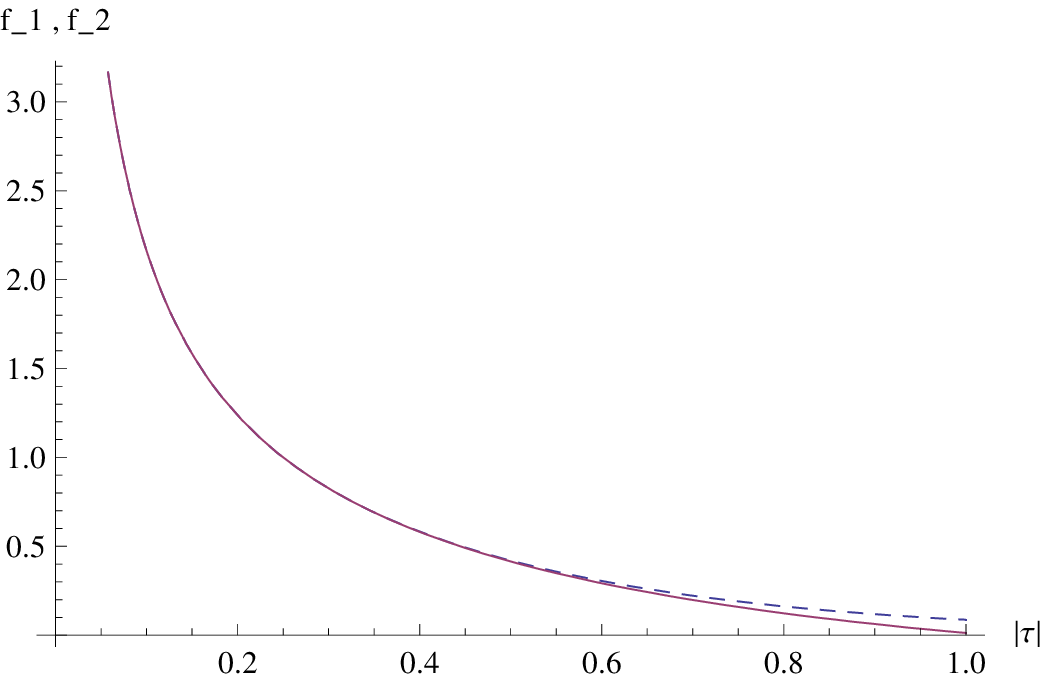}\qquad
\qquad
\includegraphics[height=4.5cm]{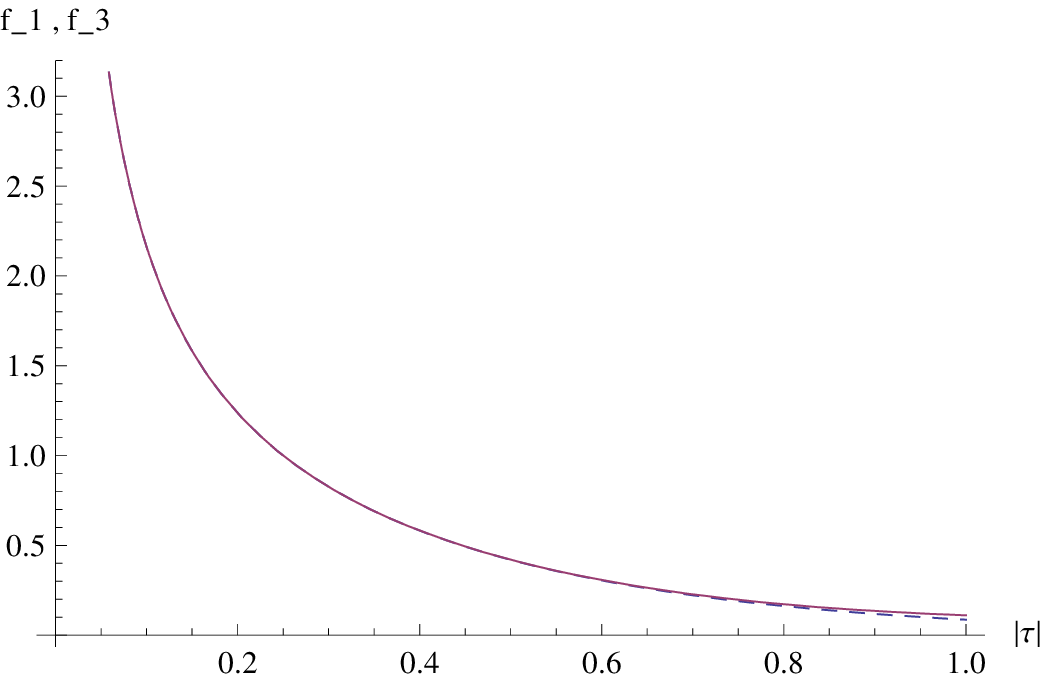}}
\end{figure}
\begin{figure}[htbp]
\centerline{\includegraphics[height=4.5cm]{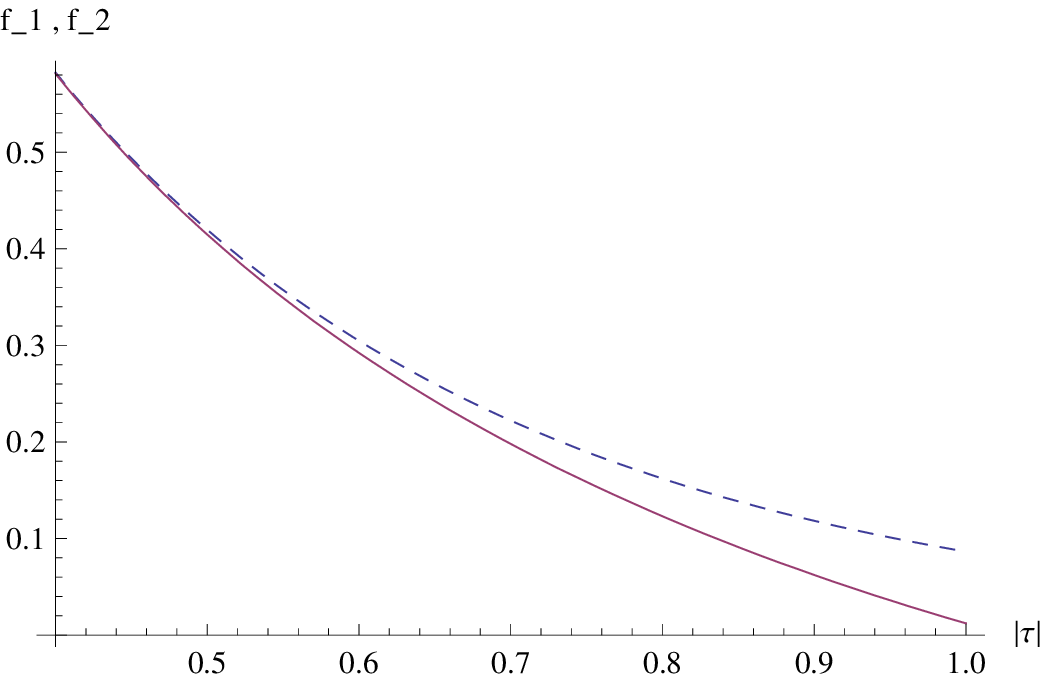}\qquad
\qquad
\includegraphics[height=4.5cm]{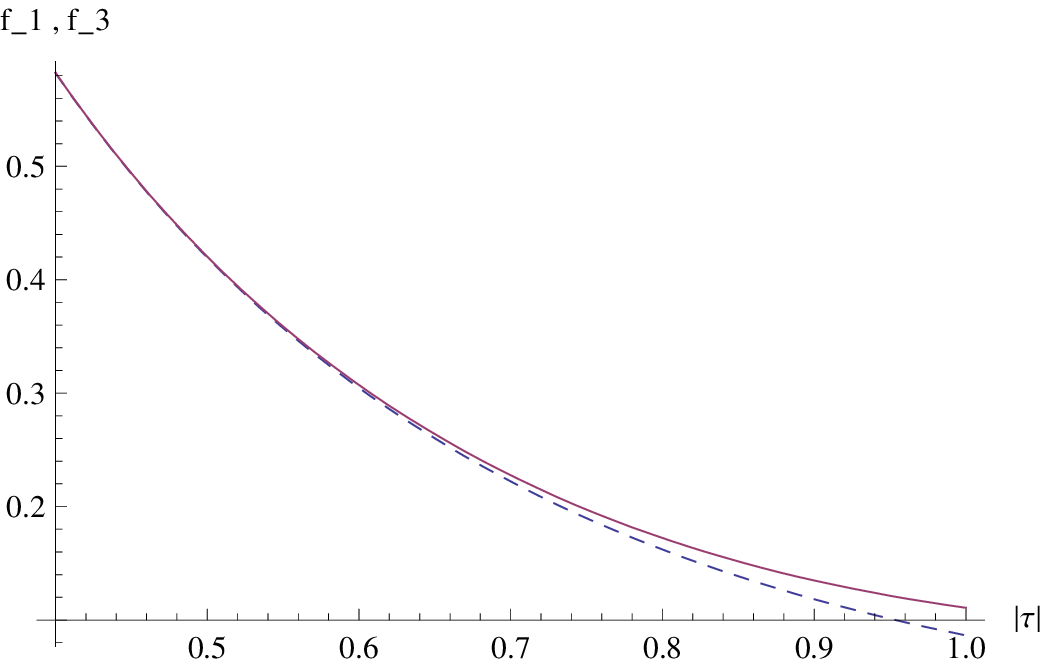}}\caption {
Plot of: (left) $f_1(|\tau|)$ (continuous line) and $f_2(|\tau|)$ (dashed line), and (right) $f_1(|\tau|)$ (continuous line) and $f_3(|\tau|)$ (dashed line).}
\end{figure}
\newpage
\par
We plan to follow this work by extending these computations to the
kink sector of the model. The idea is to compute the one-loop kink
mass shift in the framework developed in Reference \cite{AMAJWMJM}
using Dirichlet boundary conditions instead of the periodic boundary
conditions that are more conventional in quantum field theory . It
will also be of great interest to perform the same program using
more general families of boundary conditions, combining the method
developed in \cite{gen,AMAJWMJM} with the formalism developed in
references \cite{mdj1,mmj,mj}.

\end{document}